\documentclass[letter]{aa}
\usepackage[dvips]{graphicx}
\usepackage{txfonts}
\def\HII{H{\sc ii} }

\def\HC{HC~H{\sc ii}}

\def\kms{\mbox{km~s$^{-1}$}}

\begin{document}

\title{The hyperyoung \HII\ region in G24.78+0.08 A1\thanks{Based on observations obtained
at the Very Large Array (VLA) of the National Radio Astronomy Observatory (NRAO). NRAO is a facility of the national Science
Foundation operated under cooperative agreement by Associated Universities,
Inc.}}

\author{M.\ T.\ Beltr\'an \inst{1} \and R. Cesaroni \inst{2}  \and L.\
Moscadelli \inst{2} \and C.\ Codella \inst{3}} 

\offprints{M. T. Beltr\'an, \email{mbeltran@am.ub.es}}

\institute{
Departament d'Astronomia i Meteorologia, Universitat de Barcelona, Mart\'{\i} i Franqu\`es 1, 08028 Barcelona, Catalunya, Spain  
\and
INAF, Osservatorio Astrofisico di Arcetri, Largo E.\ Fermi 5, 50125 Firenze, Italy
\and
INAF, Istituto di Radioastronomia, Sezione di Firenze, Largo E.\ Fermi 5, 50125 Firenze, Italy}

\date{Received date; accepted date}

\titlerunning{The hyperyoung \HII\ region in G24.78+0.08}
\authorrunning{Beltr\'an et al.}

\abstract
{G24.78+0.08~A1 is a $20 M_\odot$ star surrounded by a hypercompact (HC) \HII\
region, driving a CO bipolar outflow, and located at the center of a
massive rotating toroid undergoing infall towards the \HC\ region. Recent water maser
observations suggest that the \HC\ region is expanding and accretion onto
the star is halted.}
{This study aims to confirm the expansion scenario proposed for the
\HC\ region on the basis of recent H$_2$O maser observations.}
{We carried out continuum VLA observations at  1.3~cm and 7~mm with the A
array plus Pie Town configuration to map the \HC\ region towards G24.78+0.08~A1.}
{The emission of the \HC\ region has been resolved and shows a ring shape structure.
The profiles of the emission obtained by taking slices at different angles
passing through the barycenter of the  \HC\ region confirm the shell
structure of the emission. The ratio between the inner and the outer radius
of the shell, $R_{\rm i}/R_{\rm o}$, derived fitting the normalized
brightness temperature profile passing through the peak of the 7~mm emission,
is 0.9, which indicates that the shell is thin. The deconvolved outer radius
estimated from the fit is 590~AU. These results imply that the \HC\ region in
G24~A1 cannot be described in terms of a classical, homogeneous \HII\
region but is instead an ionized
shell. This gives support to the model of an expanding wind-driven, ionized
shell suggested by the kinematics and distribution of the H$_2$O masers
associated with the \HC\ region. According to this model, the \HC\
region is expanding on very short times scales, 21--66~yr.}
{}

\maketitle

\section{Introduction}

In the context of massive star formation, significant progress has been recently
made with the detection of Keplerian circumstellar disks around newly born
B-type stars, and of massive toroids around young O-type stars. These results
lend support to the accretion models that describe the formation of high-mass
stars as a scaled-up version of the formation of solar-type stars. Among the
objects where rotation evidence has been inferred, one stands unique: this is
G24.78+0.08 A1 (hereafter G24~A1). G24 A1 is one of the three massive rotating
toroids detected in G24.78+0.08 (Beltr\'an et al.~\cite{beltran04},
\cite{beltran05}), a massive 
star formation region located at a distance of 7.7~kpc. At the center of G24 A1,
an unresolved hypercompact (HC) \HII\  region has been detected at 1.3~cm
by Codella et al.~(\cite{codella97}). According to these authors, the
spectral type of the star is at least O9.5, based on the free-free continuum
flux, implying a stellar mass $\sim 20 M_\odot$.
Very Large Array (VLA) observations in the NH$_3(2,2)$ line have revealed
that the gas in the toroid is undergoing infall towards the \HC\ region
(Beltr\'an et al.~\cite{beltran06}), suggesting that accretion onto the star
might still be going on, even through the \HC\ region as described in some
models (see e.g.  Keto \& Wood \cite{keto06}). However, recent VLBA proper
motion measurements of H$_2$O masers associated with the \HC\ region
(Moscadelli et al. \cite{moscadelli07}; hereafter MGCBF) indicate that the
latter might be expanding, thus questioning the possibility of accretion onto
the star.

To discriminate between the accretion and expansion scenarios, knowledge of
the internal structure of the \HC\ region is crucial. According to MGCBF, the
expansion should be driven by the stellar wind and the \HC\ region should
thus have a shell-like appearance. With this in mind, we have imaged the
free-free emission of the \HC\ region in G24~A1 in order to resolve its
structure.

\section{VLA Observations}
\label{obs}

We carried out interferometric observations in the 1.3~cm and 7~mm continuum
with the VLA on April 7 and 8, 2006. The observations were carried out with the
A array plus Pie Town configuration. The field phase center was set to the
position $\alpha$(J2000) = $18^{\rm h}   36^{\rm h} 12\fs66$, $\delta$(J2000) =
$-07\degr 12' 10\farcs15$. The detection of strong H$_2$O and CH$_3$OH maser
emission, respectively at 1.3~cm and 7~mm, within a few arcsec from the
continuum source allowed us to self-calibrate on the maser spots and then apply
the phase corrections to the continuum. The absolute position of the continuum
at 1.3~cm was shifted 17~mas eastward and 35~mas northward to allow matching of
our H$_2$O maser spot positions with those measured with much greater accuracy
by MGCBF with the VLBA. Final maps were created with the ROBUST parameter set
equal to 0 at 1.3~cm, and with natural weighting at 7~mm. The resolution and
sensitivity (3$\sigma$) attained were $0\farcs095\times0\farcs056$ (PA=$8\degr$)
and 0.25~mJy~beam$^{-1}$ at 1.3~cm, and $0\farcs061\times0\farcs047$
(PA=$-4\degr$) and 0.86~mJy~beam$^{-1}$ at 7~mm. Continuum images at 21 and 6~cm
were obtained from VLA archive data (projects AB544 and AB515). The flux density
at 6~cm and the upper limit at 21~cm were calculated  from gaussian fits to
G24~A1 and B (Codella et al.~\cite{codella97}), made with the task JMFIT of AIPS.
Since the angular resolutions (5\farcs1$\times$3\farcs6 and
5\farcs5$\times$4\farcs2 respectively at 6 and 21~cm) are insufficient to
resolve A1 from B (separation $\sim$4\arcsec), in the fit we fixed the diameter
of A1 to the half power beam width (HPBW) of the corresponding map, and its
position to that obtained from our VLA + Pie~Town observations. All parameters
of G24~B were left free.

\begin{table}
\caption[] {Parameters of the 1.3~cm and 7~mm radio continuum of the \HC\ region
in G24~A1}
\label{par_cm}
\begin{tabular}{ccccccc}
\hline
\multicolumn{2}{c}{Position$^a$} \\ 
 \cline{1-2} 
\multicolumn{1}{c}{$\alpha({\rm J2000})$} &
\multicolumn{1}{c}{$\delta({\rm J2000})$} &
\multicolumn{1}{c}{$\lambda$} &
\multicolumn{1}{c}{$T_{\rm sb}$} &
\multicolumn{1}{c}{$S_\nu$} &
\multicolumn{1}{c}{$R^{b}$} 
\\
\multicolumn{1}{c}{h m s}&
\multicolumn{1}{c}{$\degr$ $\arcmin$ $\arcsec$} &
\multicolumn{1}{c}{(cm)} &
\multicolumn{1}{c}{(K)} & 
\multicolumn{1}{c}{(mJy)} &
\multicolumn{1}{c}{(AU)}
\\
\hline
18 36 12.555  & $-07$ 12 10.83 &1.3 &$4.8\times 10^3$ & 72 &580 \\
18 36 12.556  & $-07$ 12 10.80 &0.7 &$2.4\times 10^3$ &101 &550 \\ 
\hline 

\end{tabular}

(a) obtained from a 2D gaussian fitting \\ 
(b) geometrical mean of the major and minor semi-axes, obtained from gaussian
deconvolution \\

\end{table}

\begin{figure}
\centerline{\includegraphics[angle=0,width=8.5cm]{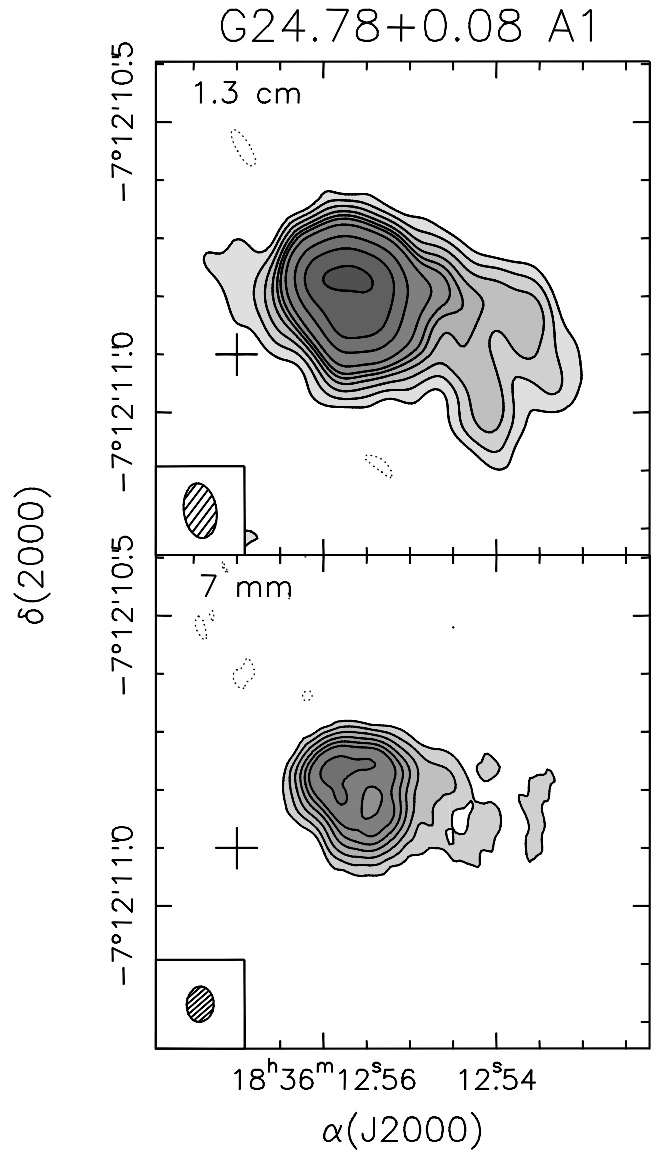}}
\caption{
Map of the VLA 1.3~cm ({\it top panel}) and 7~mm ({\it bottom panel}) continuum
emission towards  G24~A1. The contour levels are $-3$, 3, 6, 10, 15, 20, 25, 35, 60,
90, and 120 times $\sigma$, where $\sigma$ is 0.08 mJy beam$^{-1}$ at 1.3~cm,
and 0.29 mJy beam$^{-1}$ at 7~mm. The synthesized beams are drawn in the bottom left
corner.  The black cross indicates the position of the 1.4~mm continuum source
(Beltr\'an et al.\ \cite{beltran04}).} 
\label{cont} 
\end{figure}

\begin{figure}
\centering
\centerline{\includegraphics[angle=-90,width=8.5cm]{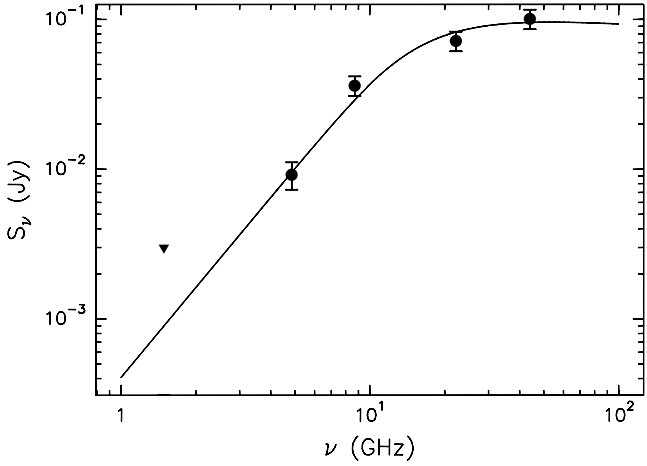}}
\caption{
Radio continuum spectrum of G24~A1. Values at 21 and 6~cm are from VLA data
archive, at 3.6~cm from Forster \& Caswell (\cite{forster00}), and at 1.3~cm
and 7~mm from this study. The triangle at 21~cm indicates an upper limit. The maximum brightness temperature measured in 
the synthesized beam is
 $\sim5\times10^3$~K at 1.3~cm and $\sim~2\times~10^3$~K at 7~mm, indicating that the
 emission is rather optically thick.}
\label{classic}
\end{figure}

\section{Results and discussion}

\subsection{Structure of the \HC\ region}
\label{uchii}

Figure~\ref{cont} shows the emission of the \HC\ region towards G24 A1 at 1.3~cm and 7~mm
wavelengths. Table~\ref{par_cm} reports the absolute position, the  maximum brightness
temperature measured in the synthesized beam, $T_{\rm sb}$, the flux density, $S_\nu$, and
the deconvolved radius, $R$, of the \HC\ region at both wavelengths. In Fig.~\ref{classic} we
show the radio continuum spectrum of G24~A1 and the fit obtained for a
classical Str\"omgren \HII\ region with radius of 0\farcs13 (1000~AU at 7.7~kpc)
and a stellar Lyman continuum of $6.7\times10^{47}$~s$^{-1}$, corresponding
to an O9.5 type star.

The \HC\ region appears to be almost circular at both wavelengths, despite a
faint ``tail'' extending towards the southwest. This slight asymmetry could
be due to lower density of the ambient gas, as suggested by the fact that the
peak of the 1.3~mm continuum emission imaged by Beltr\'an et al.
(\cite{beltran04}) lies on the opposite side of the \HC\ region (see cross in
Fi.g~\ref{cont}).  Such a ``tail'' is unlikely to be related to the  velocity
gradient observed on a larger scale in the CH$_3$CN lines (Beltr\'an et
al.~\cite{beltran04}), as the directions of the two are quite different
(PA$\simeq$245\degr\ and 222\degr\ respectively for the ``tail'' and the
gradient). Moreover, the expansion of the \HC\ region cannot affect the
velocity field over a region 10 times larger, like that where the gradient is
observed.

\begin{figure}
\centerline{\includegraphics[angle=-90,width=8.3cm]{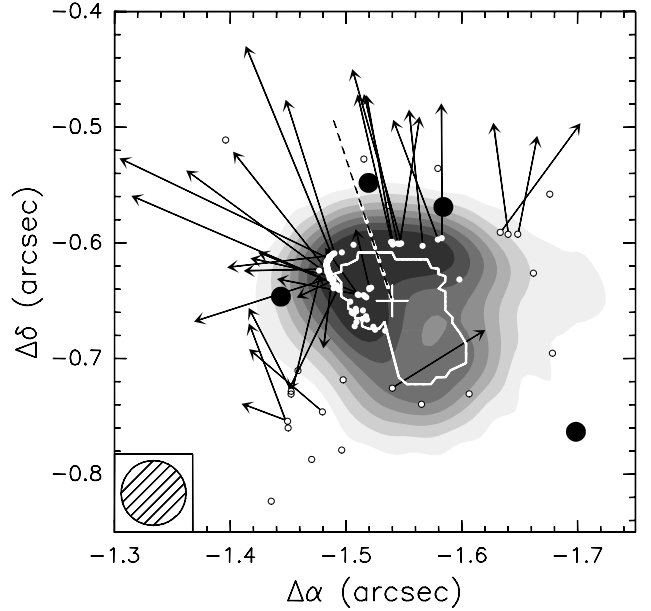}}
\caption{
Map of the VLA 7~mm towards G24~A1 ($greyscale$) created with ROBUST  equal
to 0 and reconstructed with a circular beam of $0\farcs056 \times
0\farcs056$. The contour levels range from 1.3 to 9.1 in steps of
1.3~mJy/beam. The positions of 22.2~GHz water and 6.7~GHz methanol maser
spots (from MGCBF) are denoted respectively
by white and black circles. Black arrows show the measured absolute proper
motions of the water maser spots (also from MGCBF). The white cross marks the
position of the barycenter of the 7~mm continuum map. The white line connects the
peaks of the profiles obtained from a large number of slices through the
barycenter (one every 10\degr\ in PA).  The dashed line indicates the
direction along which the normalized brightness temperature profile in
Fig.~\ref{bright} has been computed.}
\label{ring}
\end{figure}

As can be seen in Figure~\ref{cont}, the emission is resolved at both
wavelengths, and shows limb brightening, especially in the eastern edge of
the \HC\ region. The resolution at 7~mm is high enough to resolve the
structure of the \HC\ region, which resembles a ring. To better study this
ring morphology, a 7~mm map with ROBUST=0 and reconstructed with a circular
beam of $0\farcs056 \times 0\farcs056$ was created. Using this map, we
obtained profiles of the emission at different angles by taking slices
passing through the barycenter of the \HC\ region (cross in Fig.~\ref{ring}),
located at $\alpha$(J2000) = $18^{\rm h}   36^{\rm h} 12\fs557$,
$\delta$(J2000) = $-07\degr 12' 10\farcs80$. Such emission profiles, not shown here, show two
peaks, as one would expect in the case of an emission ring. In
Fig.~\ref{ring}, we have outlined the ring by joining the positions of these
peaks.  This suggests that what we are observing in the continuum is an
ionized gas shell. This is not in contradiction with the continuum spectrum
of the region (Fig.~\ref{classic}), because as discussed by Shull
(\cite{shull80}), the free-free spectrum of a shell \HII\ region is basically
indistinguishable from that of a homogeneous, spherical Str\"omgren
\HII\ region with the same angular radius and Lyman continuum.  The only
distinction between the shell and the spherical models is the presence of
limb brightening at optically thin frequencies, as observed in G24~A1.

\begin{figure}
\centering
\centerline{\includegraphics[angle=-90,width=8.4cm]{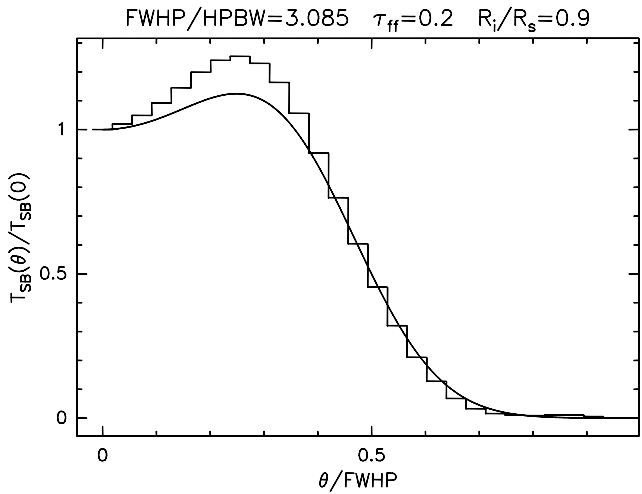}}
\caption{Normalized brightness temperature profile (histogram) along the
dashed line in Fig.~\ref{bright}. The offset is normalized with respect to
the measured FWHP. The curve represents the best fit obtained with a shell
model of the \HII\ region -- see text.}
\label{bright}
\end{figure}

Although the existence of an emission shell appears evident, the resolution
of our observations is not sufficient to derive the intrinsic diameter of the
ring in Fig.~\ref{ring}. For this purpose we have taken the brightness
temperature profile along a cut (dashed line in Fig.~\ref{ring}) passing
through the barycenter of the 7~mm image and the peak of the emission.
This profile (see Fig.~\ref{bright}) has been normalized both in temperature
and radius, dividing the former by the brightness temperature measured at the
position of the barycenter ($T_{\rm B}(0)\simeq1800$~K) and the latter by the
full width at half power (FWHP=0\farcs17) of the \HC\ region.  For a shell
\HII\ region, this profile depends on three parameters only: the ratio
FWHP/HPBW, the optical depth $\tau_{\rm ff}$ at the center of the
\HII\ region, and the ratio $R_{\rm i}/R_{\rm o}$ between the inner and outer
radius of the shell.  In our case, we know that FWHP/HPBW=3.085 and
$\tau_{\rm ff} = -\ln(1-T_{\rm B}(0)/T_{\rm e}) \simeq 0.2$ for an electronic
temperature of $10^4$~K. The best fit is then obtained for
$R_{\rm i}/R_{\rm o} \simeq 0.9$ and is shown in Fig.~\ref{bright}.

Such a result is in agreement with the prediction of Shull~(\cite{shull80})
that the shell should be very thin, i.e. $(R_{\rm i}/R_{\rm o}-1)\ll 1$. From
the fit it is also possible to estimate the deconvolved radius of the shell,
which is 590~AU. Note that this is significantly smaller than the value of
$\sim 1000$~AU needed to fit the observed radio continuum spectrum of the
\HC\ region (see Fig.~\ref{classic}).  This is due to the fact that the
\HC\ region is not perfectly spherically symmetric, but shows an emission
``tail'' towards the southwest (see above), which in spite of its
relatively low brightness, conveys a non-negligible fraction of the total
flux -- $\sim$28\% ($\sim$20~mJy) at 1.3~cm and $\sim$30\% ($\sim30$~mJy)
at 7~mm. This ``extra'' flux cannot be accounted for by the simple shell
model.

\subsection{Expansion of the \HC\ region}

As shown before, the \HC\ region in G24~A1 cannot be described in terms of a
classical compact and spherical \HII\ region, but of an ionized shell. This
result confirms the interpretation given by MGCBF to explain the distribution
and proper motions of the 22.2~GHz water maser features towards G24~A1. As seen
in Fig.~\ref{ring}, the water masers are well correlated with the eastern edge
of the \HC\ region and have proper motions directed approximately perpendicular
to the arc described by the maser spots. According to MGCBF, this configuration
indicates that the water masers are tracing expansion, which in turn suggests
that the \HC\ region is expanding. Because of the high velocity ($\sim 40$
km\,s$^{-1}$), such an expansion cannot be lead by the thermal pressure of the
ionized gas, as in the classical case depicted, e.g., by De Pree et
al.~(\cite{depree95}). Therefore, it must be some additional mechanism driving
the expansion. This is also suggested by the line width (40~\kms)
of the H66$\alpha$ recombination line (Goddi, private communication): this is
well in excess of the thermal width in \HII\ regions (typically 20~km\,s$^{-1}$)
and thus requires additional broadening (see Sewilo et al.~\cite{sewilo04}),
such as large scale motions, consistently with our expanding model. Note
however, that the width of the recombination line appears to indicate a smaller
expansion velocity (Robinson et al.~\cite{robinson82}; Welty \cite{welty83}) than that traced by the water masers. However, a caveat is in
order. The problem posed by the detection of such broad recombination lines is
not yet solved (see Sewilo et al.~\cite{sewilo04}), especially because their strength is
inconsistent with the \HII\ region being optically thick in the continuum. For
this reason, to describe the recombination line emission from hypercompact \HII\
regions, and thus obtain a correct estimate of the expansion velocity, one needs
a better model, where kinematics and radiative transfer are properly taken into
account, which  goes beyond the purposes of the present work.

Following the model by Shull~(\cite{shull80}), MGCBF, conclude that the \HII\
region expansion may be driven by a powerful stellar wind.  In such a model,
there is an initial phase of free expansion at the wind velocity, when the wind
sweeps up its own mass in interstellar matter. After that, the system evolves
into a four-zone structure, in which a thin, dense ionized shell containing most
of the swept up material is created. In this process two phases can be
identified:  a pressure-driven expansion, followed by a momentum-driven
expansion after a critical time that depends on the wind mechanical luminosity
and density of the surrounding environment.  For a given wind mechanical
luminosity and density, the radius ($R_{\rm HII}$) and velocity ($V_{\rm HII}$)
of the expanding shell can be expressed as a function of time
(Shull~\cite{shull80}). For our estimates we have adopted a Lyman continuum of
$6.7\times10^{47}$~s$^{-1}$ (see Sect.~\ref{uchii}), an inner-to-outer radius
ratio of 0.9 for the shell (see Sect.~\ref{uchii}), and the fiducial values of
$10^7$~cm$^{-3}$, $10^{36}$~erg~s$^{-1}$, and 2000~\kms\ respectively for the
environment density, mechanical luminosity of the wind, and wind speed (see
MGCBF for a detailed discussion on the input model parameters).

\begin{figure}
\centering
\centerline{\includegraphics[angle=0,width=8.4cm]{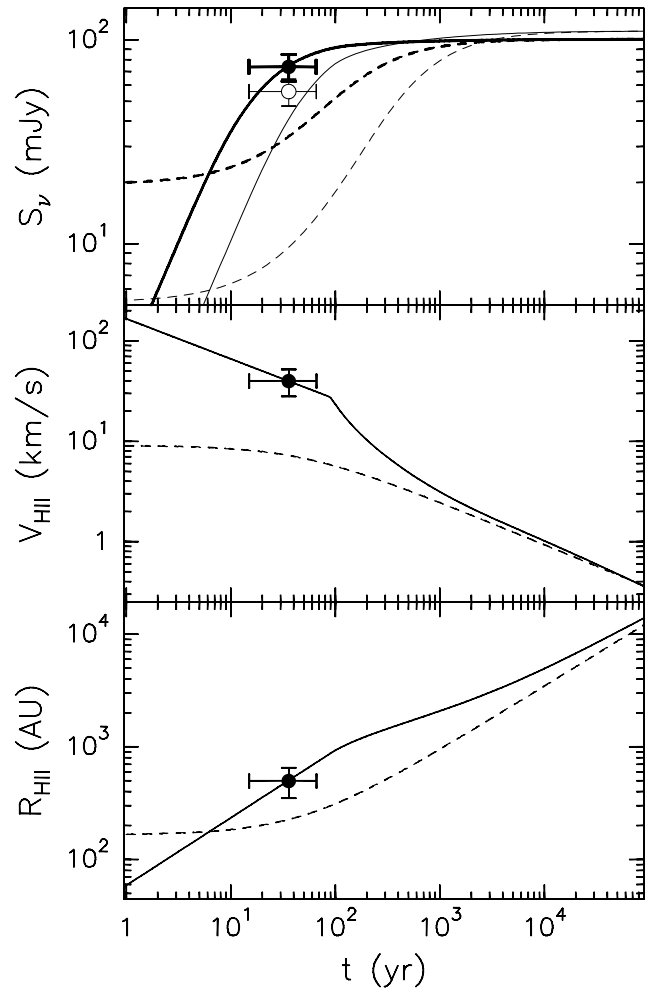}}
\caption{
Flux ({\it top panel}), velocity ({\it middle panel}), and radius ({\it
bottom panel}) of a \HC\ region as a function of time for two different
models: a classical \HII region ({\it dashed lines}), and a wind-driven shell
({\it solid lines}). The thin lines in {\it top panel} indicate the flux at
1.3~cm, and the thick lines the flux at 7~mm. The measured fluxes (56~mJy at
1.3~cm and 74~mJy at 7~mm) do not take into account the contribution from the
southwestern emission ``tail''.  The vertical error bars correspond to
uncertainties of 30\% for the radius and expansion velocity, and 15\% for the
flux. The error bar on $t$ corresponds to the uncertainty estimated by
MGCBF.}
\label{models}
\end{figure}

The values of the shell radius and velocity as a function of time are shown
in Fig.~\ref{models}, as well as the 1.3~cm and 7~mm continuum fluxes of the
\HC\ region. For the sake of comparison, we also show the same quantities
for a classical Str\"omgren \HII\ region (dashed curves). The corresponding
observed values with the associated uncertainties are also reported and
demonstrate that only the wind-driven can fit {\it all} the data. Note that
the fluxes used here (56~mJy at 1.3~cm and 74~mJy at 7~mm) do not contain the
contribution from the emission ``tail'' mentioned in Sect.~\ref{uchii}.

Given the very short time scale involved (21--66~yr), the size variation
expected in only 5~yr ($\sim$10~mas) should be easy to reveal using the
H$_2$O maser spots as probes, thus providing us with a robust test for
our model.

\section{Conclusions}

Our radio continuum images with $<$90~mas resolution have revealed a ring
shaped structure in the \HC\ region at the center of the rotating toroid
imaged by Beltr\'an et al. (\cite{beltran04}), supporting the model of a
wind-driven, ionized shell proposed by MGCBF. We thus conclude
that very likely the \HC\ region is expanding on very short time scales
(21--66~yr). This finding appears to contradict the fact that the gas on a
larger scale is infalling towards the \HC\ region (Beltr\'an et al.
\cite{beltran06}). A possible explanation is that the infalling gas is not
accreting any more onto the star, but is stopped (and possibly diverted into
the outflow observed by Furuya et al. \cite{furuya02}) at the surface of the
\HC\ region, right at the shock front traced by the H$_2$O maser spots. It is
worth noting that the CH$_3$OH masers observed by MGCBF appear
to lie further from the \HC\ region than the H$_2$O masers, suggesting that
they might be located in the pre-shock material. If this is the case,
methanol masers should be participating to the infall and their proper
motions should be directed towards the \HC\ region, unlike the water masers
which are clearly expanding from it. Future VLBI multi-epoch measurements are
bound to confirm or rule out this scenario.

\begin{acknowledgements}

MTB is supported by MEC grant AYA2005-08523-C03. 

\end{acknowledgements}

\end{document}